\documentclass[longauth,oldversion]{aa}  
\usepackage{epsfig}
\usepackage{graphicx}
\usepackage{natbib}
\usepackage{longtable}
\usepackage{txfonts}

\bibpunct{(}{)}{;}{a}{}{,}

\begin{document}
\renewcommand{\topfraction}{0.85}
\renewcommand{\bottomfraction}{0.7}
\renewcommand{\textfraction}{0.15}
\renewcommand{\floatpagefraction}{0.66}

   \title{Chandra and H.E.S.S. observations of the Supernova Remnant CTB\,37B}
\titlerunning{Chandra and H.E.S.S. observations of the SNR CTB\,37B}
\authorrunning{H.E.S.S. Collaboration}

\author{F. Aharonian\inst{1,13}
 \and A.G.~Akhperjanian \inst{2}
 \and U.~Barres de Almeida \inst{8} \thanks{supported by CAPES Foundation, Ministry of Education of Brazil}
 \and A.R.~Bazer-Bachi \inst{3}
 \and B.~Behera \inst{14}
 \and M.~Beilicke \inst{4}
 \and W.~Benbow \inst{1}
 \and K.~Bernl\"ohr \inst{1,5}
 \and C.~Boisson \inst{6}
 \and V.~Borrel \inst{3}
% ...
}

\institute{
Max-Planck-Institut f\"ur Kernphysik, P.O. Box 103980, D 69029
Heidelberg, Germany
\and
 Yerevan Physics Institute, 2 Alikhanian Brothers St., 375036 Yerevan,
Armenia
\and
Centre d'Etude Spatiale des Rayonnements, CNRS/UPS, 9 av. du Colonel Roche, BP
4346, F-31029 Toulouse Cedex 4, France
\and
Universit\"at Hamburg, Institut f\"ur Experimentalphysik, Luruper Chaussee
149, D 22761 Hamburg, Germany
\and
Institut f\"ur Physik, Humboldt-Universit\"at zu Berlin, Newtonstr. 15,
D 12489 Berlin, Germany
\and
LUTH, Observatoire de Paris, CNRS, Universit\'e Paris Diderot, 5 Place Jules Janssen, 92190 Meudon, 
France
\and
IRFU/DSM/CEA, CE Saclay, F-91191
Gif-sur-Yvette, Cedex, France
\and
University of Durham, Department of Physics, South Road, Durham DH1 3LE,
U.K.
\and
Unit for Space Physics, North-West University, Potchefstroom 2520,
    South Africa
\and
Laboratoire Leprince-Ringuet, Ecole Polytechnique, CNRS/IN2P3,
 F-91128 Palaiseau, France
\and 
Laboratoire d'Annecy-le-Vieux de Physique des Particules, CNRS/IN2P3,
9 Chemin de Bellevue - BP 110 F-74941 Annecy-le-Vieux Cedex, France
\and
Astroparticule et Cosmologie (APC), CNRS, Universite Paris 7 Denis Diderot,
10, rue Alice Domon et Leonie Duquet, F-75205 Paris Cedex 13, France
\thanks{UMR 7164 (CNRS, Universit\'e Paris VII, CEA, Observatoire de Paris)}
\and
Dublin Institute for Advanced Studies, 5 Merrion Square, Dublin 2,
Ireland
\and
Landessternwarte, Universit\"at Heidelberg, K\"onigstuhl, D 69117 Heidelberg, Germany
% ...
}

  \offprints{J.\,L.~Skilton (phy3j2ls@ast.leeds.ac.uk)}

   \date{Received; Accepted}

  \abstract {The $>$100~GeV $\gamma$-ray source, HESS\,J1713$-$381, apparently
    associated with the shell-type supernova remnant (SNR) CTB\,37B, was
    discovered using H.E.S.S. in 2006. X-ray follow-up
    observations with \textit{Chandra} were performed in 2007 with the aim of
    identifying a synchrotron counterpart to the TeV source and/or
    thermal emission from the SNR shell. These new \textit{Chandra} data,
    together with additional TeV data, allow us to investigate the
    nature of this object in much greater detail than was previously
    possible. The new X-ray data reveal thermal emission from a $\sim$4$'$ region in 
    close proximity to the radio shell of CTB\,37B. The temperature of this
    emission implies an age for the remnant of $\sim$5000 years and an ambient gas density of $\sim$0.5\,cm$^{-3}$. Both these estimates are considerably uncertain due to the asymmetry of the SNR and possible modifications of the kinematics due to efficient cosmic ray (CR) acceleration. A bright ($\approx$7\,$\times$10$^{-13}$erg\,cm$^{-2}$s$^{-1}$) and unresolved ($<$1$\arcsec$) source (CXOU\,J171405.7$-$381031) with a soft ($\Gamma\approx3.3$)
		non-thermal spectrum is also detected in coincidence with the radio shell. Absorption
    indicates a column density consistent with the thermal emission
    from the shell suggesting a genuine association rather than a
    chance alignment. The observed TeV morphology is consistent with an 
    origin in the complete shell of CTB\,37B. The lack of diffuse non-thermal
    X-ray emission suggests an origin of the $\gamma$-ray emission via the
    decay of neutral pions produced in interactions of protons and nuclei, rather
    than inverse Compton (IC) emission from relativistic electrons.
}

   \maketitle
%

\section{Introduction}

Approximately 50 sources of very-high-energy (VHE; $>$100~GeV) $\gamma$-rays have now been detected along the plane of our Galaxy \citep{HESS:Scan2,TevReview}. Some of these sources are (at least partially) coincident with supernova remnants (SNRs), consistent with the paradigm of cosmic-ray acceleration at SNR shocks. Resolved VHE $\gamma$-ray emission from the \emph{shells} of two nearby ($d$$\sim$1~kpc) SNRs has been detected from: RX\,J0852.0$-$4622  \citep{HESS:VelaJnr} and RX\,J1713.7$-$3946 \citep{HESS:rxj1713p3}. However, in more distant objects, emission from the
shell is often not resolvable with current instruments and there is ambiguity in the origin of the VHE emission. In several cases pulsar wind nebulae (PWNe; for a review see \cite{PWN:review}), powered by neutron stars left behind in the supernova responsible for the remnants, are viable alternative acceleration sites (for example in the cases of HESS\,J1813$-$178 \citep{HESS:1813}, and HESS\,J1804$-$216 \citep{HESS:Scan2}).
 
The sensitivity to detect much more distant $\gamma$-ray SNRs ($d$$\gg$1kpc), and hence learn more about the population of such objects, has been achieved only in a few regions of our galaxy where deep observations have taken place with the most sensitive available TeV instruments. One such region is that within a few degrees of RX\,J1713.7$-$3946, where a $>$50\,hour observation with the 4-telescope H.E.S.S. system results in a point-source sensitivity of $1.5\times10^{-13}$ photons~cm$^{-2}$~s$^{-1}$ above 1~TeV (or $\sim$0.7\% of the flux of the Crab Nebula). Within this region (l~$\sim$348$^{\circ}$, b~$\sim$0$^{\circ}$), a $\gamma$-ray source, HESS\,J1713$-$381 was detected in coincidence with the CTB\,37 region of radio emission \citep{HESS:Scan2}.
This region consists of three supernova remnants, CTB\,37A (which consists of two separate remnants overlapping in projection: G\,348.5+0.1 and G\,348.5+0.0) and
CTB\,37B (G\,348.7+0.3) \citep{Kassim91}.
The first deep TeV measurements resulted in a significant excess close to CTB\,37B and a less significant excess close to CTB\,37A \citep{HESS:Scan2}. Including the more recent H.E.S.S. data discussed here, the latter source (HESS\,J1714$-$385) is detected with a high level of confidence and is discussed in detail in a separate paper \citep{HESS:ctb37a}. 

At radio wavelengths CTB\,37A and B have very similar properties \citep{Clark75}: similar surface brightnesses ($\sim$0.2~Jy~arcsec$^{-2}$), spectral indices ($\alpha\sim$0.3 \citep{Kassim91}) and sizes ($\sim$5$'$). CTB\,37A is thought to be interacting with surrounding molecular clouds (with densities of 100--1000~cm$^{-3}$) \citep{Reynoso00} and appears to have significant OH maser emission, indicative of dense shock heated gas \citep{Frail96}. No maser emission has been confirmed from the direction of CTB\,37B.

The distance to CTB\,37B has been estimated by a number of authors. \cite{Caswell75} used HI absorption measurements to determine the kinematic distance to CTB\,37B (using the rotation curve of \cite{Schmidt65})
to be 10.2$\pm$3.5 kpc. A similar result was found for the distance to CTB\,37A leading to the claim that CTB\,37A and B are close not only in projection but also along the line of sight. 
Reinterpreting this result using a more recent galactic rotation curve model \citep{Brand93} yields
a distance range of 5-9~kpc (we note that the value adopted in the catalogue of \citet{Green06}
is 8~kpc).


The age of CTB\,37B is also uncertain. \cite{Clark75} give age estimates for both CTB\,37A and B of $\sim$1500~years and claim that there is a possibility that either source could be the remnant of the supernova of AD\,393. This claim is further investigated by \cite{Downes84} who concluded that although such an association is possible, the remnants may in fact be considerably older.

Both to improve our understanding of the SNR CTB\,37B and to investigate the nature of HESS\,J1713$-$381, sensitive X-ray measurements of this region were considered highly desirable. We observed the field of CTB\,37B with \textit{Chandra} in early 2007. These new X-ray data, together with additional H.E.S.S. data, are discussed here, with the aim of
elucidating the relationship between these two objects.


\section{H.E.S.S. observations and analysis}

The region toward CTB\,37 was observed between April and October 2004 using the High Energy Stereoscopic System (H.E.S.S.) during a survey of the Galactic Plane, and in pointed observations toward the SNR\,RX\,J1713.7$-$3946, yielding 37 hours of on-source live-time \citep{HESS:Scan2}. VHE gamma-ray emission was detected at a 6.3$\sigma$ significance level from a region of rms size 0.06$^{\circ}$$\pm$~0.04$^{\circ}$, spatially coincident with the SNR complex, at a position RA~=~17$^{\rm h}$13$^{\rm m}$57.6$^{\rm s}$, Dec~=~38$^{\circ}$12$\arcmin$0$\arcsec$ (epoch J2000) and was announced as a new source HESS\,J1713$-$381. Initial spectral analysis of this source led to a photon index of 2.27~$\pm$~0.48 and a flux above 200 GeV of (4.2~$\pm$~1.5)~$\times$10$^{-12}$cm$^{-2}$s$^{-1}$. The region has since been re-observed with H.E.S.S. and the data set used for analysis here includes all telescope pointings within 1.8$^{\circ}$ from the best fit position of the source, leading to a total on-source live-time of 55.4 hours. Correcting this to the equivalent on-axis exposure yields 40.8 hours. The observations were performed over a large range of zenith angles (14$^{\circ}$-60$^{\circ}$) with an average value of 35$^{\circ}$.

Analysis was carried out using standard H.E.S.S. procedures (for details see \cite{HESS:crab}). An independent $\emph{model}$ analysis \citep{deNaurois06} was performed and yielded consistent results. Hard cuts (image size cut 200 photoelectrons (p.e.)) provide a higher signal to noise ratio and a better PSF for 2-D images at the expense of an increased energy threshold and were applied for morphological analysis. Standard cuts (80 p.e.) provide a wider energy range and were thus applied for spectral extraction. The average energy threshold of the data set was 370 GeV for hard cuts and 250 GeV for standard cuts.

Two different background estimation models were used in the analysis of the data (described in detail in \cite{Berge07}). For spectral analysis, a reflected-region background model was applied using only off-regions with angular displacement less than 1.8$^{\circ}$ from the best fit source position in an attempt to reduce systematic errors arising from the strong exposure gradient across the field of view (due to most pointed observations being directed at RX~J1713.7$-$3946). For morphological analysis and 2-D image generation, used for determining the best fit position of the source, a ring background model was used with ring radius 0.5$^{\circ}$, excluding all known $\gamma$-ray sources such as HESS\,J1714$-$385 from the background region.

The integration region used for estimating the flux and differential energy spectrum of the VHE emission was a circle of radius 0.187$^{\circ}$ centered on the best fit position. This size was chosen so as to maximise the encompassed excess whilst minimising contamination from HESS\,J1714$-$385. From the source-fitting it was estimated that this region contained 93\% of the total emission from HESS\,J1713$-$381 and included an estimated 11\% contamination from HESS\,J1714$-$385.

\begin{figure}
\centering
\epsfig{file=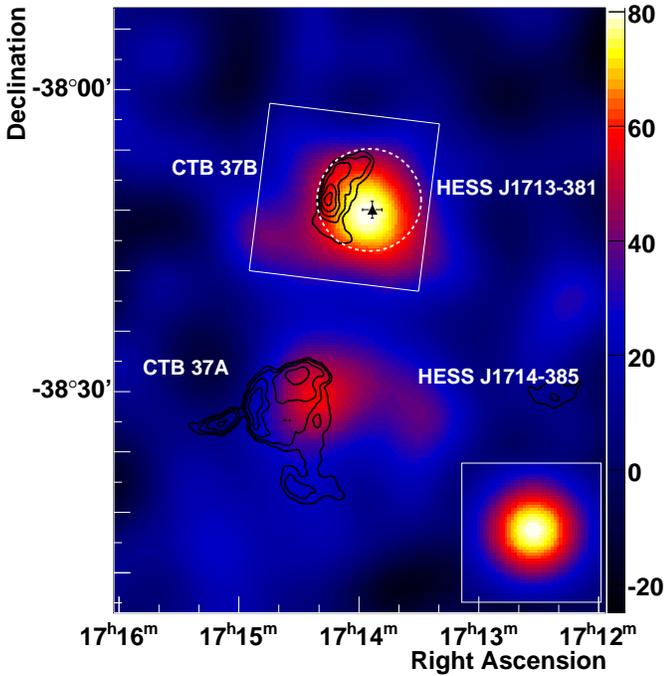, width=1\linewidth, clip=} \\
\caption{Acceptance corrected $\gamma$-ray excess map of the SNR complex CTB\,37 after Gaussian smoothing with a radius $\sigma$$_{\rm sm}$ of 2.9$\arcmin$. The black contours represent 0.08, 0.16, 0.32, 0.64, 1.28 and 2.56 Jy/beam radio emission from the Molonglo Galactic Plane Survey (beam size $\sim$1$\arcmin$) \citep{Green99}. The field of view of the \textit{Chandra} observation is shown by a white box surrounding CTB\,37B. The estimated size of the supernova remnant, centered on the position stated by \cite{Green06}, is illustrated by a dashed white circle. The simulated PSF is shown in the box in the bottom right corner and is smoothed in the same way as the excess map. The colour scale of the map is in units of counts per 2$\pi$$\sigma$$_{\rm sm}^2$~$\equiv$~counts per 0.0145 degrees$^2$.}
\label{fig:Gskymap}
\end{figure}

Figure~\ref{fig:Gskymap} shows a smoothed, background subtracted and acceptance corrected image of the region surrounding CTB\,37. TeV emission is observed from both HESS\,J1713$-$381 (the centroid of which is marked with a filled black triangle) and HESS\,J1714$-$385. Radio contours, taken from the Molonglo Galactic Plane Survey at 843~MHz \citep{Green99}, show emission only from the Eastern side of the SNR CTB\,37B as well as a faint plateau of emission extending to the South-East of the remnant. A circle of radius 5.1$\arcmin$ has been drawn onto the image in Figure~\ref{fig:Gskymap} to represent the size of the SNR based on the partial radio shell. The centroid of the TeV emission from HESS\,J1713$-$381 is located at RA~=~17$^{\rm h}$13$^{\rm m}$54$^{\rm s}$~$\pm$~4$^{\rm s}$$_{stat}$, Dec~=~$-$38$^{\circ}$11$\arcmin$58$\arcsec$~$\pm$~45$\arcsec$$_{stat}$, close to the centre of this proposed shell. The systematic error in the source location is estimated to be 20$\arcsec$ in both coordinates. The source shows an excess of 292~$\pm$~36 counts with a statistical significance (using the likelihood method of \cite{li_ma83}) of 8.6$\sigma$. A somewhat higher significance is obtained using the \emph{model} analysis \citep{HESS:ctb37a}.

\begin{figure}
\centering
\epsfig{file=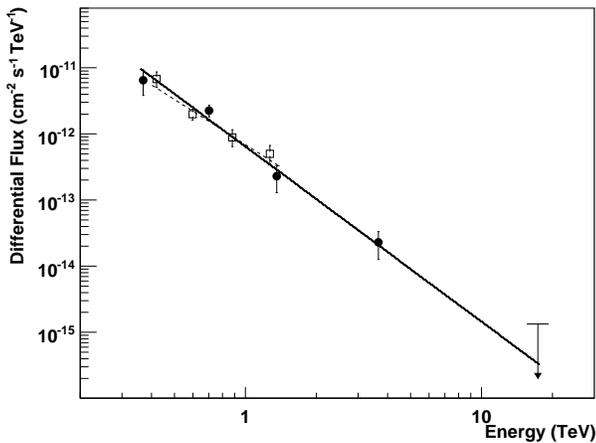, width=1\linewidth, clip=} \\
\caption{Differential energy spectrum of HESS\,J1713$-$381. Solid symbols are derived from the full data set and are fitted with a power law with photon index $\Gamma$~=~2.65~$\pm$~0.19 (solid line). A 90\% confidence upper limit is given at energies above 10~TeV. The empty squares and the dashed line represent the data acquired during previous observations \citep{HESS:Scan2}.}
\label{fig:Gspec}
\end{figure}

A power law with photon index $\Gamma$~=~2.65~$\pm$~0.19$_{stat}$~$\pm$~0.20$_{sys}$ and a differential normalization at 1~TeV of (6.5~$\pm$~1.1$_{stat}$~$\pm$~1.3$_{sys}$)\,$\times$10$^{-13}$cm$^{-2}$s$^{-1}$\,TeV$^{-1}$ provides a satisfactory description of the data ($\chi^2$/dof~=~3.8/3) as shown in Figure~\ref{fig:Gspec}. The integral flux above 200~GeV was found to be (5.6~$\pm$~1.5$_{stat}$~$\pm$1.1$_{sys}$)\,$\times$10$^{-12}$cm$^{-2}$s$^{-1}$ and the energy flux in the range 0.5-5~TeV is $\approx$2$\times$10$^{-12}$erg~cm$^{-2}$s$^{-1}$.

A joint morphological fit of HESS\,J1714$-$385 and HESS\,J1713$-$381 was made with a model consisting of two Gaussian emission profiles convolved with the PSF of the instrument. The intrinsic rms width of HESS\,J1713$-$381 extracted from this fit, 2.6$\arcmin$~$\pm$~0.8$\arcmin$, is compatible with the size found during the H.E.S.S. survey \citep{HESS:Scan2}. An unresolved thin shell of radius $r$ has an equivalent Gaussian rms of $\sigma$~$\simeq$0.6$r$. For a filled sphere $\sigma$~$\simeq$0.45$r$. The measured width is consistent with a shell of radius $\sim$4$-$6$\arcmin$, compatible with the estimated radio shell size.

\section{X-ray observations and analysis}

The CTB\,37B region was observed with the Advanced CCD Imaging Spectrometer (ACIS) of the \textit{Chandra} X-ray Observatory on the 2$^{\rm nd}$ February 2007 for 26\,ks (observation ID~6692). The raw data were analysed using the Chandra Interactive Analysis of Observations (CIAO version~3.4, CALDB version~3.4.1). The data set was unaffected by soft proton flares enabling the full observation time to be utilised for analysis purposes.

\begin{figure}
\centering
\epsfig{file=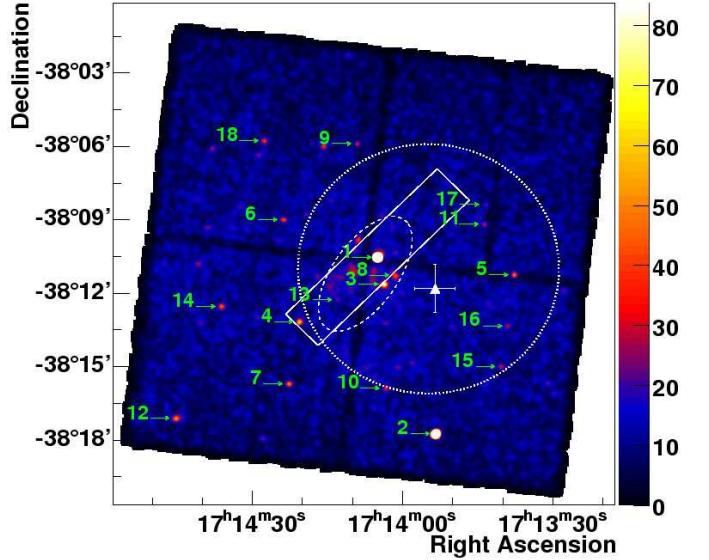, width=1\linewidth, clip=} \\
\caption{X-ray image (1-5~keV) of the \textit{Chandra} field of view showing the positions of the 18 detected point sources on the four ACIS CCD chips. The image is a count map after Gaussian smoothing with $\sigma$$_{\rm sm}$~=~5$\arcsec$. Source~\#1 is located very close to one of the CCD chip edges. The position of the centroid of the TeV emission from HESS\,J1713-381 is marked by a white triangle. The white box shows the region from which the emission profile in Figure~\ref{fig:Xslice} is taken and the dashed white ellipse shows the region from which the diffuse X-ray spectrum was extracted. The colour scale shows counts per 2$\pi$$\sigma$$_{\rm sm}^2$ and has been truncated to better show the fainter point sources in contrast to the background. The maximum prior to truncation was 1214 at the position of source~\#1. This truncation forces the brighter sources to appear larger than their true size. As in Figure~\ref{fig:Gskymap}, the dotted circle illustrates the estimated size of the SNR~CTB\,37B}
\label{fig:Xrawmap}
\end{figure}

Figure~\ref{fig:Xrawmap} shows a count map of the region after Gaussian smoothing with a $5''$ radius to match the off-axis PSF. This image illustrates the positions of 18 point-like sources, described in Table~\ref{tab:xray}, detected at a $>$5$\sigma$ level within the \textit{Chandra} field of view using the CIAO \emph{wavdetect} algorithm. The most significant of these sources, source~\#1, is located 3$\arcmin$ North-East of the centroid of the TeV source HESS\,J1713-381. 

\begin{figure}
\centering
\epsfig{file=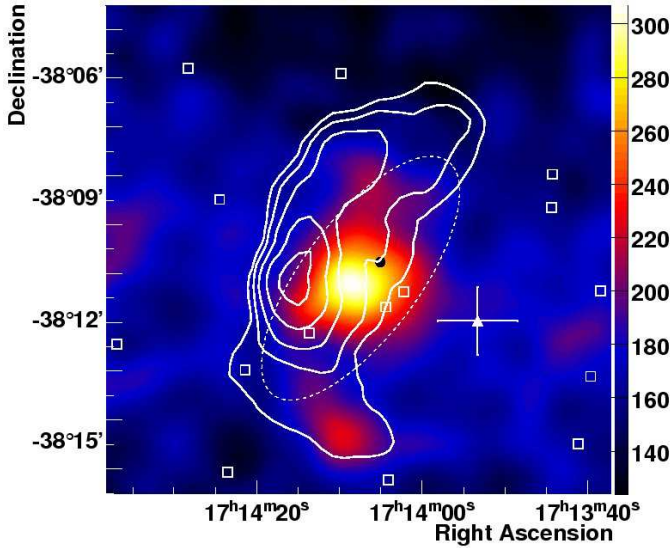, width=1\linewidth, clip=} \\
\caption{Exposure corrected count map (1-5~keV) after subtraction of emission from the 18 point sources detected in the field of view. The image has been Gaussian smoothed with $\sigma$$_{\rm sm}$ = 50$\arcsec$ to highlight the diffuse emission. The colour scale is in units of counts per 2$\pi$$\sigma$$_{\rm sm}^2$. The black circle shows the position of source \#1. White square symbols show the location of the other point sources which have been removed from the image. The white contours represent the same radio contours as in Figure~\ref{fig:Gskymap}. The position of the centroid of HESS\,J1713$-$381 is shown as a white triangle.}
\label{fig:Xsmoothmap}
\end{figure}

Figure~\ref{fig:Xsmoothmap} shows an exposure corrected, smoothed excess map after removal of emission from the point sources described in Table~\ref{tab:xray}. Extended diffuse emission is observed at a $>$11$\sigma$ level to the East of HESS\,J1713$-$381 and appears to be contained within the region defined by the radio contours. The diffuse emission extends over all four ACIS CCD chips. This peak is spatially coincident with the radio shell but is not aligned with the peak of the radio emission and is instead offset to the West by $\sim$1.6$'$. This shift is highly significant considering the 0.6$\arcsec$ astrometric accuracy of \textit{Chandra}. No evidence for significant temperature variations across the diffuse emission was found and spectral properties on each individual ACIS chip were consistent. Hence the region used to extract the reported diffuse spectrum extends over all four chips, and is indicated by a dashed white ellipse in Figures~\ref{fig:Xrawmap} and~\ref{fig:Xsmoothmap}.

\begin{figure}
\centering
\epsfig{file=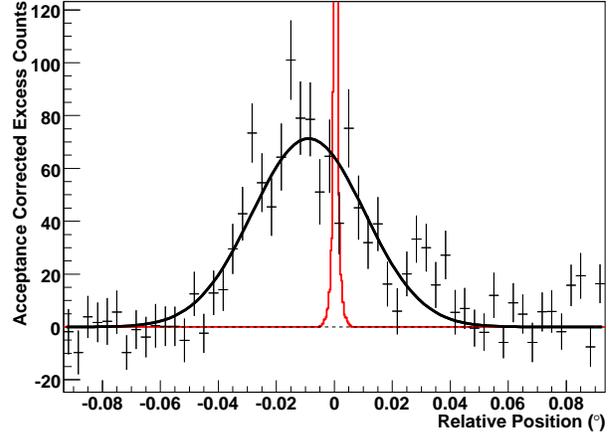, width=1\linewidth, clip=} \\
\caption{Acceptance corrected emission profile taken from the rectangular region shown in Figure~\ref{fig:Xrawmap}, after the removal of emission from the 18 detected point sources. This region is centered on the best fit position of source~\#1. A Gaussian fit to the data is shown by the smooth curve. The simulated PSF model for (the subtracted) source \#1 is shown in red. Source~\#1 is $\sim$10 times brighter ($\sim$1200 counts) than the observed diffuse emission. The background is estimated from regions of equal size and shape displaced to the north and south of the diffuse emission.}
\label{fig:Xslice}
\end{figure}

Figure~\ref{fig:Xslice} shows a profile of the X-ray emission in the rectangular region shown in Figure~\ref{fig:Xrawmap}, after the removal of the 18 point sources found within the \textit{Chandra} field of view. This region is centered on the position of the X-ray source CXOU\,J171405.7$-$381031 (source~\#1). The simulated PSF model for source \#1 is shown in red. The rms width of the diffuse emission extracted from a Gaussian fit to the data (black curve, $\chi^2$/dof~=~86.6/57) is $\sigma$~=~1.2$\arcmin$~$\pm$~0.1$\arcmin$.

\begin{figure}
\centering
\epsfig{file=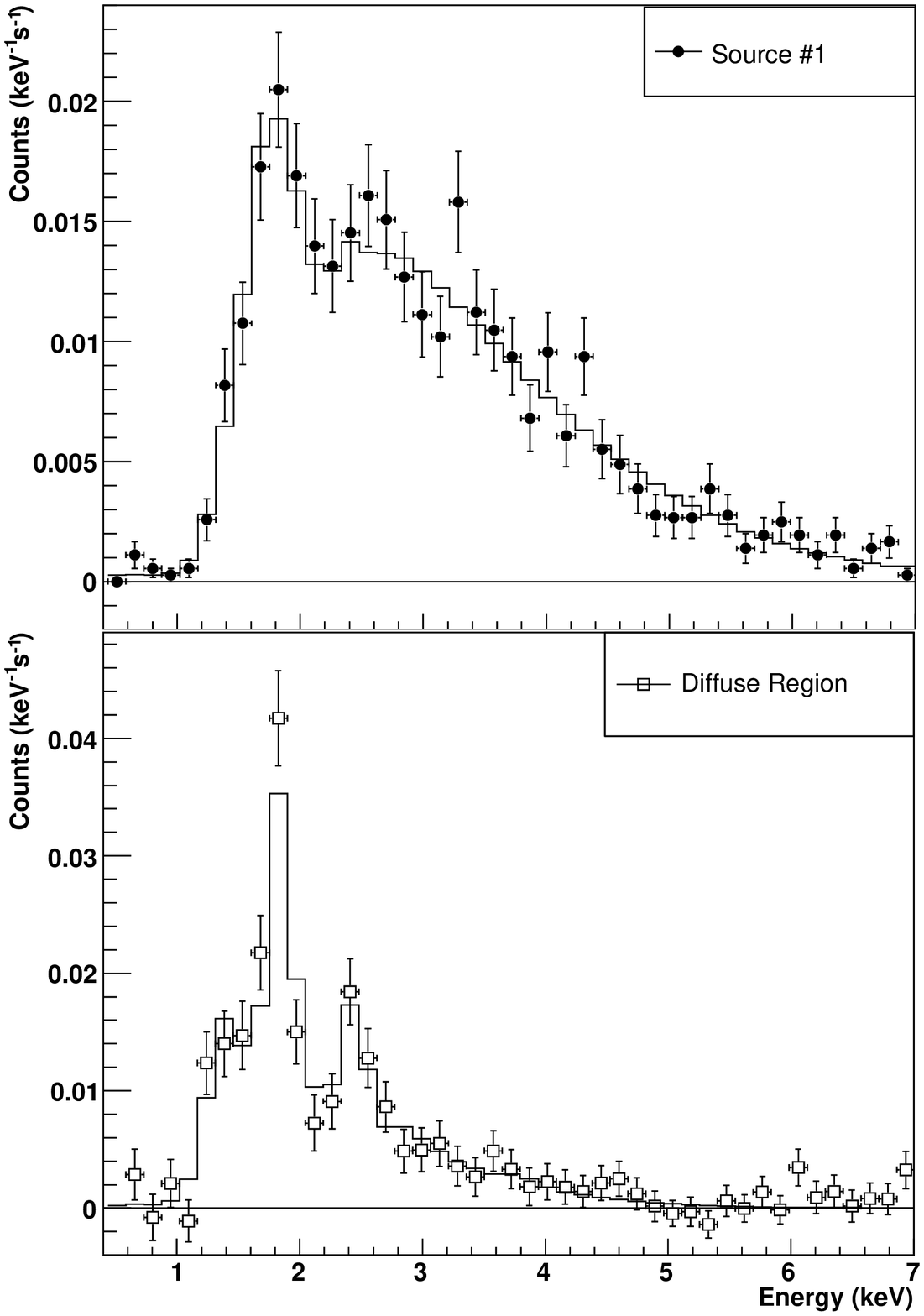, width=1\linewidth, clip=} \\
\caption{\textit{Top:} Count rate versus energy for source~\#1 (black filled symbols) compared to the best fit model for an absorbed power law. \textit{Bottom:} Count rate versus energy for the diffuse emission compared to the best fit model for an absorbed NEI model. Fit parameters are described in the text.}
\label{fig:Xspectrum}
\end{figure}

The energy spectrum of source~\#1 was taken from an integration region of radius $\sim$1.7$\arcsec$ centered on the best fit position. This spectrum is well described by a single absorbed power law with N$_{\rm H}$~=~(4.1$^{+0.3}_{-0.2}$)\,$\times$10$^{22}$~cm$^{-2}$ and $\Gamma$~=~3.3$^{+0.2}_{-0.1}$ with $\chi^2$/dof~=~68.5/63 (the probability of a blackbody model fit to this source was $\sim$0.2\%). Both sources \#2 and \#3 are consistent with blackbody models with kT\,$<$0.5~keV. The other 15 point sources were too weak to allow a meaningful spectral analysis to be performed.

Regions around the 18 detected X-ray point sources were excluded from the data set before extracting the spectrum of the extended diffuse emission. The background was estimated from a large region covering all 4 ACIS chips and excluding regions around the point sources and the diffuse emission. The resulting spectrum was best fit with an absorbed non-equilibrium ionisation (NEI) model with $\tau$~=~(7$^{+9}_{-3}$)\,$\times$10$^{10}$s~cm$^{-3}$ ($\chi^2$/dof~=~68.9/62) and is shown in the bottom panel of Figure~\ref{fig:Xspectrum}. A marginally acceptable fit is also possible with $\tau$ set to the maximum allowed value of $5\,\times\,10^{13}$ s\,cm$^{-3}$, indicating that the source is approaching equilibrium. No evidence was found for diffuse emission above 5~keV. Parameters of fits with different spectral models are given in Table~\ref{tab:spectra}. The consistent column densities from the model fits suggest a source~\#1 and the diffuse emission lie at approximately the same distance.

\begin{table*}
\centering
      \begin{tabular}{c c c c r r }
\hline\hline
ID & Name & RA & Dec & Counts & Significance \\
 & CXOU\,J & 17$^{\rm h}$ & -38$^{\circ}$ &  & $\sigma~~~~$\\
\hline
1 & 171405.7-381031 & 14$^{\rm m}$5.758$^{\rm s}$ & 10$\arcmin$31.32$\arcsec$ & 1171 & 158.9 \\
2 & 171354.1-381844 & 13$^{\rm m}$54.129$^{\rm s}$ & 18$\arcmin$44.24$\arcsec$ & 445 & 38.6 \\
3 & 171404.3-381237 & 14$^{\rm m}$4.385$^{\rm s}$ & 12$\arcmin$37.51$\arcsec$ & 64 & 23.2 \\
4 & 171421.3-381310 & 14$^{\rm m}$21.38$^{\rm s}$ & 13$\arcmin$10.19$\arcsec$ & 55 & 19.6 \\
5 & 171438.4-381114 & 14$^{\rm m}$38.43$^{\rm s}$ & 11$\arcmin$14.2$\arcsec$ & 40 & 11.4 \\
6 & 171323.4-381641 & 13$^{\rm m}$23.45$^{\rm s}$ &16$\arcmin$41.8$\arcsec$ & 40 & 9.8 \\
7 & 171424.4-380959 & 14$^{\rm m}$24.49$^{\rm s}$ & 9$\arcmin$59.1$\arcsec$ & 28 & 9.1 \\
8 & 171402.2-381115 & 14$^{\rm m}$2.224$^{\rm s}$ & 11$\arcmin$15.4$\arcsec$ & 19 & 8.4 \\
9 & 171409.7-380652 & 14$^{\rm m}$9.77$^{\rm s}$ & 6$\arcmin$52.7$\arcsec$ & 23 & 8.2 \\
10 & 171404.0-381652 & 14$^{\rm m}$4.09$^{\rm s}$ & 16$\arcmin$52.4$\arcsec$ & 23 & 7.6 \\
11 & 171444.3-380910 & 14$^{\rm m}$44.35$^{\rm s}$ & 9$\arcmin$10.5$\arcsec$ & 19 & 6.9 \\
12 & 171346.0-381705 & 13$^{\rm m}$46.08$^{\rm s}$ & 17$\arcmin$5.8$\arcsec$ & 37 & 6.5 \\
13 & 171413.6-381215 & 14$^{\rm m}$13.66$^{\rm s}$ & 12$\arcmin$15.7$\arcsec$ & 14 & 6.4 \\
14 & 171436.8-381331 & 14$^{\rm m}$36.88$^{\rm s}$ & 13$\arcmin$31.8$\arcsec$ & 31 & 6.3 \\
15 & 171341.0-381558 & 13$^{\rm m}$41.07$^{\rm s}$ & 15$\arcmin$58.7$\arcsec$ & 19 & 5.9 \\
16 & 171339.6-381320 & 13$^{\rm m}$39.61$^{\rm s}$ & 13$\arcmin$20.0$\arcsec$ & 15 & 5.4 \\
17 & 171344.2-380822 & 13$^{\rm m}$44.22$^{\rm s}$ & 8$\arcmin$22.0$\arcsec$ & 13 & 5.1 \\
18 & 171428.2-380545 & 14$^{\rm m}$28.27$^{\rm s}$ & 5$\arcmin$45.6$\arcsec$ & 23 & 5.1 \\
\hline
     \end{tabular}
      \caption{Properties of all point sources detected above a 5$\sigma$ level in the \textit{Chandra} field of view using the CIAO \emph{wavdetect} algorithm. Statistical errors on the positions are $\sim$3$\arcsec$ for the weakest sources on each axis. The astrometric accuracy of \textit{Chandra} is 0.6$\arcsec$}
\label{tab:xray}
  \end{table*}
  
\begin{table*}
\centering
      \begin{tabular}{l c c c c c c c }
\hline\hline
Element & Model & \multicolumn{4}{c}{Best Fit Parameters} & Flux (1-5 keV)& $\chi^2$/dof \\
 & (with absorption)& N$_{\rm H}$ (10$^{22}$ cm$^{-2}$) & $\Gamma$ & kT (keV) & $\tau$ (10$^{10}$s~cm$^{-3}$) & (10$^{-13}$erg~cm$^{-2}$s$^{-1}$) & \\
\hline
Source~\#1 & Power law & 4.1$^{+0.3}_{-0.2}$ & 3.3$^{+0.2}_{-0.1}$ & & & 6.7 & 68.5/63\\
Source~\#1 & Blackbody & 2.1 $\pm$ 0.2 & & 0.76 $\pm$ 0.03 & & 5.9 & 100.2/63\\
Source~\#1 & Two-temperature & 2.9$^{+0.2}_{-0.3}$  & & 0.52$^{+0.06}_{-0.04}$ & & 6.4 & 68.0/61\\
 & Blackbody & & & 1.6$^{+0.6}_{-0.3}$ & & \\
Diffuse emission & NEI & 3.9$^{+0.3}_{-0.2}$ & & 0.8 $\pm$ 0.1 & 7$^{+9}_{-3}$ & 2.9 & 68.8/62 \\
Diffuse emission & NEI & 3.8$^{+0.5}_{-0.4}$ & & 0.7 $\pm$ 0.1 & 5$\times$10$^3$ (fixed) & 2.7 & 75.5/63 \\
Diffuse emission & Power law & 4.2$^{+0.6}_{-0.5}$ & 5.3$^{+0.6}_{-0.5}$ & & & 2.8 & 101.6/63\\
\hline
     \end{tabular}
      \caption{Best fit spectral parameters for the point source CXOU\,J171405.7$-$381031 and the diffuse  X-ray emission with various models. Errors are given at the 1$\sigma$ level. The fluxes listed are the absorbed fluxes}
\label{tab:spectra}
  \end{table*}
  
Timing analysis of the source was difficult due to its position on the chip edge leading to an artificial modulation at 10$^{-3}$Hz caused by instrumental dither effects, and the 3.2~s frame time. However, we find no evidence for variability on 3$\times$10$^{-3}$ to 0.2~Hz timescales.

\section{Discussion}

\subsection{The X-ray emission of the SNR CTB\,37B}

The X-ray measurements presented here allow a better assessment of the
age and environment of CTB\,37B than was previously possible.
From the age and distance of the remnant an external density can be
estimated. Unfortunately the distance to the remnant is rather
uncertain. Here we adopt a 7~kpc distance based on the HI absorption measurements of \cite{Caswell75} scaled to more recent galactic rotation measurements. Assuming
that the thermal X-ray diffuse emission is indeed associated with the
SNR and that electrons and ions are in thermal equilibrium, the temperature obtained from the NEI fit ($kT =$
0.8~$\pm$~0.1 keV) can be used to estimate the SNR age using the
Sedov-Taylor phase equations 
\citep[e.g.][]{Borkowski01}. The temperature of the thermal emission implies a shock velocity, $v_s\sim$800\,$(T/0.8\rm keV)^{1/2}(\mu/0.62m_{\rm H})^{-1/2}$\,km\,s$^{-1}$ and the age of the remnant is estimated to be $\sim$4900~(d/7\,kpc)($\mu$/0.62\,m$_{\rm H}$)$^{1/2}$($T$/0.8\,keV)$^{-1/2}$~years (where $\mu$ is the mean mass per particle). At a distance of 7~kpc the estimated radius of ~5$\arcmin$ corresponds to an intrinsic size close to 10~pc.
A spherical Sedov expansion then implies an external density of $\sim$0.8 (E$_k$/10$^{51}$~erg)\,cm$^{-3}$ where E$_k$ is the kinetic energy of the SN explosion. An age of 1600~years, compatible with an origin in the supernova of AD 393, would imply $n\sim$0.1~(E$_k$/10$^{51}$~erg)\,cm$^{-3}$.

There are three main complications which could affect these estimates. Firstly, the Sedov solution holds only approximately, as the region surrounding CTB\,37B appears to be inhomogeneous, leading to an asymmetric expansion. Secondly, the temperature equilibrium between electrons and ions assumed above is not seen in many young SNRs, see \cite{Rakowski05} for a recent review. In the case of non-equilibration, the velocity estimate must be increased by a factor $\sqrt{(1 + T_i/T_e)/2}$ (asuming a pure hydrogen gas), making  the remnant younger by the same factor. Finally, efficient acceleration of cosmic rays at the shock would result in less heating of the thermal gas for a given shock velocity (because a fraction of the dissipated kinetic energy goes into the non-thermal population). The more efficient the acceleration, the faster the shock has to be to achieve a given gas temperature, leading to an over estimate of the SNR age using the conventional Sedov-Taylor gas dynamic solution.  In the approximation of a steady-state, plane-parallel shock \citep[see e.g.][]{Drury81}, it may be shown that if the fraction of the downstream enthalpy in accelerated particles is $\vartheta$, then the velocity estimate must be increased by a factor of approximately $1 / \sqrt{1-\vartheta}$, and the age  estimate decreased accordingly. For example, the combined effect of non-equilibrium temperatures, $T_i\,=\,3T_e$, and very efficient CR acceleration, $\vartheta =$0.4, would change the estimated age of the remnant to $\sim$2700\,years.

The thermal X-ray flux can be used as a cross-check for this scenario.
Due to absorption the luminosity implied by the measured flux is only a small fraction of the radiated power. The de-absorbed bolometric flux, $f_{bol}$, is $\approx$2.6$\times$10$^{-11}$~erg\,cm$^{-2}$\,s$^{-1}$. At 0.8\,keV ($\sim$10$^7$K) the cooling coefficient, $\Lambda$ is $\approx$2.7$\times$10$^{-23}$~erg\,cm$^{3}$\,s$^{-1}$ \citep{Sutherland93}. The ion density in the emission region is connected to $f_{bol}$ via $f_{bol} = \Lambda V\,n^2\,/\,4\pi\,d^2$, assuming thermal equilibrium and equal number densities for electrons and ions. Taking the volume, $V$, to be that of an ellipsoid with principle axes equal to the FWHM of the X-ray emision along its major axis, $a$\,=\,2.8$'$ (as shown in Figure~\ref{fig:Xslice}), the FWHM in the perpendicular direction, $b$\,=\,1.9$'$, and the FWHM along the line of sight (estimated as $f \times a$), we find $n$\,$\sim$$2\,(f\,d/7\,\mathrm{kpc})^{-1/2}$\,cm$^{-3}$. The non-observation of thermal X-ray emission in other parts of the shell implies $n \lesssim 1$ cm$^{-3}$ in these regions. If this emission occurs in a shock compressed region the external density is likely a factor
$\sim$4--7 lower and hence broadly consistent with the previous estimate.
A general picture therefore emerges of CTB 37\,B as a few 1000~year
old SNR expanding into a (somewhat inhomogeneous) medium with average
density $n\sim$0.5 cm$^{-3}$. This discussion assumes that the bulk of the thermal X-ray emission originates in the shocked surrounding medium; if a significant fraction were to originate in shocked SN ejecta, the above estimates would have to be substantially revised.

\subsection{X-ray point sources}

None of the 18 detected X-ray point sources were found to be spatially coincident with the centroid of the observed VHE emission and there is no obvious candidate for a compact central object of the SNR CTB\,37B. For the three brightest sources (\#1, \#2, and \#3) there are sufficient statistics for spectral analysis to be attempted and all are apparently point-like in nature. Source~\#1 (CXOU\,J171405.7$-$381031) is by far the brightest and lies in coincidence with the radio shell of CTB\,37B and close to the maximum of the diffuse X-ray emission of the shell. As it has a measured N$_{\rm H}$ consistent with that of the SNR shell, it seems plausible that there is a physical association of these two objects. In this context, source~\#1 may represent the X-ray emission of an isolated neutron star. We find no clear counterpart for this object at optical/infrared wavelengths; the closest catalogued source is the star GSC\,2.3\,S8PL008678 from the Guide Star Catalogue (Version 2.3.2) \citep{GST:23}, offset by $\sim$4$\arcsec$. The closest 8$\mu$m GLIMPSE \citep{Glimpse} source apparent by visual inspection lies $\sim$3$\arcsec$ away.


Whilst the luminosity of source \#1 ($\sim$$4\times10^{33}$ erg s$^{-1}$) lies in a range fairly typical for the X-ray nebulae of \emph{Vela-like} PWN (for example PSR\,J1823$-$13 and PSR\,J1811$-$1925, see \cite{Kargaltsev07}), the lack of extended, non-thermal emission around source~\#1 argues strongly against the PWN hypothesis.
In addition, an identification of source \#1 with the neutron star from the SN explosion of CTB\,37B would imply a large proper motion of this object. Assuming a birth place of the pulsar at the geometric centre of the remnant (RA~=~17$^{\rm h}$13$^{\rm m}$55$^{\rm s}$, Dec~=~$-$38$^{\circ}$11$'$0$''$) \citep{Green06} and an age $t$ yields an implied `kick' velocity of $\sim$$1000 (d/7\,\mathrm{kpc})(t/5000\,\mathrm{years})^{-1}$~km~s$^{-1}$. This estimate lies towards the upper edge of the range of velocities observed for pulsars (10\,--\,$\sim$2000\,km\,s$^{-1}$) \citep{Iben96,Chatterjee02,Arzoumanian02}\footnote{We note that both the X-ray and radio emission suggest a denser medium on the Eastern side of the remnant implying that the explosion center may have been significantly further East than the current geometrical center, hence reducing the required proper motion of the pulsar}. A pulsar with such a velocity is likely to be associated with a bow-shock morphology pulsar wind nebula, as seen for example in G\,359.23$-$0.82 (``the Mouse'') \citep{Gaensler04} where the pulsar has a velocity of $\approx$600\,kms$^{-1}$. No such feature is detected in this case.

Source~\#2 is spatially coincident with the star USNO$-$A2.0\,0450$-$26795004 and has a soft thermal spectrum ($kT\sim0.2$  keV) and relatively low N$_{\rm H}$ of $<$\,6$\times$10$^{21}$\,cm$^{-2}$, consistent with emission from a nearby star. 
Source~\#3 also has a low temperature ($\sim$0.4~keV) black-body spectrum but the closest catalogued star is $\sim$3$\arcsec$ away. This source is spatially coincident with the diffuse X-ray emission, and has a consistent column density (($6\pm2)\,\times$10$^{22}$cm$^{-2}$) suggesting that these objects may be physically associated.
The closest source to the centroid of VHE emission is source~\#8 but the 19 excess counts are not sufficient for spectral analysis. This source has no catalogued stellar counterpart.

\subsection{The nature of the $\gamma$-ray source HESS\,J1713$-$381}

The position and size of HESS\,J1713$-$381 make an association with
CTB\,37B very likely. However, as in the case of other objects such as
HESS\,J1813$-$178 \citep{HESS:1813}, there is possible ambiguity between 
an origin of the TeV emission in the SNR shell and a pulsar wind nebula 
interpretation. Here we discuss these alternative explanations in turn.

\subsubsection{TeV emission from a PWN}

There are now $>$10 associations of young pulsars and TeV
$\gamma$-ray sources.  These associations suggest a typical fraction
of $\sim$1\% for the conversion of rotational energy into TeV emission
\citep{Carrigan07}. Assuming that the explosion of CTB\,37B left behind a
pulsar, the TeV emission could be explained as 
inverse Compton emission from ``relic'' electrons accelerated soon 
after the pulsar birth. Assuming IC scattering only on the cosmic microwave background (CMB), the synchrotron cooling time of electrons 
responsible for TeV emission is $\sim$8000 (B/10$\,\mu\mathrm{G})^{-2}
(E_{\gamma}/1\,\mathrm{TeV})^{-1/2}$ years, long enough that electrons
injected immediately after the pulsar birth would continue to contribute to
the TeV emission. In contrast, $\sim$3~keV synchrotron emission is
associated with higher energy ($\sim$80$
(B/10\,\mu\mathrm{G})^{-1/2}\,\mathrm{TeV}$) electrons which cool somewhat
faster:
$t_\mathrm{sync}\approx\,1500\,(B/10\,\mu\mathrm{G})^{-3/2}(\epsilon_{\mathrm{sync}}/3\,\mathrm{keV})^{-1/2}$
years, and would hence be confined to a region close to the present
pulsar position.  The prototype of the class of TeV bright PWN, HESS\,J1825$-$137
\citep{HESS:1825p2}, is associated with the $\sim$2$\times10^{4}$
year old, high-spindown luminosity ($\dot{E}\approx~3\times$10$^{36}$~erg~s$^{-1}$) pulsar
PSR\,B1823$-$13.  HESS\,J1713$-$381 has similar spectral shape ($\Gamma=2.65\pm0.19$ versus
$\Gamma\approx2.4$) but is apparently somewhat smaller (diameter 10
pc cf 40~pc), younger (5~kyr cf 20~kyrs) and dimmer (0.5-5~TeV luminosity $\sim$1.2$\times$10$^{34}$\,erg\,s$^{-1}$ cf $\sim$8$\times$10$^{34}$\,erg\,s$^{-1}$).

However, even if CXOU\,J171405.7$-$381031 is indeed a high
kick velocity pulsar, the absence of extended non-thermal X-ray emission is very difficult to 
explain in this scenario. An explanation of HESS\,J1713$-$381 by
inverse Compton emission from ultra relativistic electrons in the wake
of the pulsar therefore seems rather implausible. 

\subsubsection{TeV emission from the SNR shell}

Whilst there is a clear offset of the centroid of HESS\,J1713$-$381
from the thermal X-ray and radio emission of the shell, the TeV
centroid is consistent with the estimated geometrical centre of the
remnant.  In addition, the observed size of the TeV source is
consistent with emission in a thin shell of radius $\sim$4.5$'$, consistent
with the radius of the radio shell.  If the observed radio and X-ray
emission from the Eastern part of the shell is due to a local density
and/or magnetic field enhancement, then it seems plausible that
particle acceleration may also be taking place in the Western part.
In a scenario where the TeV emission originates from inverse Compton
(IC) scattering of $>$ TeV electrons, the emission is expected to
directly trace the distribution of such electrons (assuming uniform
radiation fields on the scale of the SNR). However, X-ray synchrotron
emission is expected from the same population of electrons with a flux
$\sim F_{\mathrm{TeV}} (B/3\mu\mathrm{G})^{2} /
(U_{\mathrm{rad}}/U_{\mathrm{CMB}}$), assuming that there is no
cut-off in the electron energy spectrum between the range probed by
the TeV IC emission and that probed by $2-10$~keV synchrotron emission (see above).
The lack of measured non-thermal X-ray emission in the shell implies
an X-ray synchrotron flux $F_{\mathrm{sync}}$ significantly lower than
the observed thermal flux of $\sim$3$\,\times~10^{-13}$~erg~cm$^{-2}$~s$^{-1}$. Given the measured TeV flux of 2\,$\times~10^{-12}$~erg~cm$^{-2}$~s$^{-1}$ (0.5$-$5~TeV), this IC scenario
implies either an implausibly low B-field of $\sim$1~$\mu$G or a rather
sharp cut-off in the electron spectrum around $\sim$40~TeV.

A more natural explanation may be an origin of the TeV emission via
the decay of neutral pions produced in proton-proton interactions in
the region of the SNR shell. Following \citet{Aharonian94} the expected
$\gamma$-ray flux is $F(>$1\,TeV)\,$\sim10^{-12}\,\theta\,(E_k/10^{51}\mathrm{erg})~(d/7\,\mathrm{kpc})^{-2} (n/0.5\,\mathrm{cm}^{-3})$ cm$^{-2}$ s$^{-1}$ where $\theta$ is the CR acceleration efficiency. The measured flux
above 1~TeV (3.9$\pm$0.7)~$\times10^{-13}$ cm$^{-2}$ s$^{-1}$ implies efficient CR acceleration. The
relatively soft measured TeV spectrum ($\Gamma$ = 2.65 $\pm$ 0.19)
is somewhat surprising in this scenario but could be interpreted as
the consequence of a cut-off in the proton spectrum at energies
somewhat lower than seen for example in RX\,J1713.7$-$3946
($E_{\mathrm{max}}\sim$100\,TeV) \citep{HESS:rxj1713p3}.

\subsubsection{Association with the EGRET source 3EG\,J1714$-$3857}

HESS\,J1713-381 lies close to the 99\% confidence contour of the $>$100 MeV source 3EG\,J1714$-$3857 \citep{EGRET3} and so this object could be considered as a potential counterpart to the observed TeV excess. The SNR RX\,J1713.7$-$3946 has also been suggested as a counterpart to this emission \citep{Butt02}, as has the newly discovered $\gamma$-ray source HESS\,J1714-385 \citep{HESS:ctb37a}. Despite exhibiting much weaker TeV emission than RX\,J1713.7$-$3946, HESS\,J1713-381 may be considered as a viable counterpart to 3EG\,J1714$-$3857 due to its softer spectral shape ($\Gamma\approx2.7$). An extrapolation of the TeV spectrum down to GeV energies is marginally consistent with the measured flux of 3EG\,J1714$-$3857. However, such a spectral match is not unlikely for a chance pairing of H.E.S.S. and EGRET sources, given the respective instrumental sensitivities and typical spectral shapes \citep{Funk08}. The angular resolution and flux sensitivity of \textit{GLAST} \citep{Glast} should allow this issue to be resolved in the near future.


\section{Summary}

The new X-ray and $\gamma$-ray data presented here represent a major
improvement in our knowledge of the CTB\,37B/HESS\,J1713$-$381 system.
The discovery of thermal X-ray emission from the shell allows the
uncertainties on the age and ambient density of the remnant to be
considerably reduced, with likely values of $\sim$5000~years and
$\sim$0.5~cm$^{-3}$ emerging. The non-thermal point-source
CXOU~J171405.7$-$381031, which
is detected embedded in the diffuse thermal emission, may be a neutron star
associated with CTB\,37B, but considerable difficulties exist with this interpretation. 
It seems likely that the TeV emission originates in the shell of CTB\,37B.
Whilst an IC origin of the TeV emission cannot be excluded,
the decay of neutral pions produced
in proton-proton interactions in the (entire) shell of the supernova 
remnant seems to be the most natural explanation.

\begin{acknowledgements}

The support of the Namibian authorities and of the University of Namibia facilitating the construction and operation of H.E.S.S. is gratefully acknowledged, as is the support by the German Ministry for Education and Research (BMBF), the Max Planck Society, the French Ministry for Research, the CNRS-IN2P3 and the Astroparticle Interdisciplinary Programme of the CNRS, the U.K. Science and Technology Facilities Council (STFC), the IPNP of the Charles University, the Polish Ministry of Science and Higher Education, the South African Department of Science and Technology and National Research Foundation, and by the University of Namibia. We appreciate the excellent work of the technical support staff in Berlin, Durham, Hamburg, Heidelberg, Palaiseau, Paris, Saclay, and in Namibia in the construction and operation of the equipment. We would like to thank Ryoko Nakamura for helpful discussions regarding the X-ray analysis.

\end{acknowledgements}

\bibliographystyle{aa}
\bibliography{CTB37B}{}

\end{document}